\documentclass[a4paper,twoside,onecolumn,notitlepage]{article}%
\usepackage{amsfonts}
\usepackage{amsmath}
\usepackage{amssymb}
\usepackage{graphicx}
\usepackage{appendix}%
\providecommand{\U}[1]{\protect\rule{.1in}{.1in}}
\newtheorem{theorem}{Theorem}

\newtheorem{definition}[theorem]{Definition}
\newtheorem{example}[theorem]{Example}

\newtheorem{lemma}[theorem]{Lemma}

\newtheorem{proposition}[theorem]{Proposition}
\newtheorem{remark}[theorem]{Remark}

\newenvironment{pf}[1][Proof]{\textbf{#1.} }{\ \rule{0.5em}{0.5em}}

\newenvironment{Conclu}[1][Conclusion]{\noindent\textbf{#1.} }{\ \rule{0.0em}{0.0em}}
\begin{document}

\title{Population Dynamics with Infinite Leslie Matrices: Finite Time Properties}
\author{Jo\~{a}o F. Alves\thanks{jalves@math.ist.utl.pt}, Ant\'{o}nio
Bravo\thanks{abravo@math.ist.utl.pt} and Henrique M.
Oliveira\thanks{holiv@math.ist.utl.pt (corresponding author)}\\$^{\ast\text{,}\ddag}$Centro de An\'{a}lise Matem\'{a}tica Geometria e
Sistemas Din\^{a}micos\\$^{\dag}$Centro de An\'{a}lise Funcional e Aplica\c{c}\~{o}es\\Math. Dep., T\'{e}cnico Lisboa, Universidade de Lisboa\\Av. Rovisco Pais, 1049-001 Lisbon, Portugal}
\maketitle

\begin{abstract}
Infinite Leslie matrices, introduced by Demetrius forty years ago are
mathematical models of age-structured populations defined by a countable
infinite number of age classes. This article is concerned with determining
solutions of the discrete dynamical system in finite time. We address this
problem by appealing to the concept of kneading matrices and kneading
determinants. Our analysis is applicable not only to populations models, but
to models of self-reproducing machines and self-reproducing computer programs.
The dynamics of theses systems can also be described in terms of infinite
Leslie matrices.



\end{abstract}
\date{}

\section{Introduction}

Almost seventy years ago Paul Holt Leslie introduced in \cite{Leslie1945} and
\cite{Leslie1948}\ matrices to study age-structured populations with given
birth ratios and the population divided by age classes. Leslie introduced a
non-negative square $p\times p$ matrix $\mathbf{L}$%
\[
\mathbf{L}=\left(
\begin{array}
[c]{cccccc}%
a_{0} & a_{1} & a_{2} & \cdots & a_{p-2} & a_{p-1}\\
b_{1} & 0 & 0 & \cdots & 0 & 0\\
0 & b_{2} & 0 & \cdots & 0 & 0\\
\vdots & \vdots & \vdots & \ddots & \vdots & \vdots\\
0 & 0 & 0 & \cdots & 0 & 0\\
0 & 0 & 0 & \cdots & b_{p-1} & 0
\end{array}
\right)  ,
\]
to study the dynamics of an age-structured population. The population is
divided in age classes $l=0,1,\ldots p-1$. We consider the population vector
$\overline{\mathbf{u}}=\left(  u_{0},u_{1},\ldots,u_{p-1}\right)  $ where the
components are the number of individuals in each class. The $b_{j}$ are the
transition probabilities from age class $j-1$ to age class $j$ and the $a_{l}$
are the mean birth rates of each individual in the class $l$.

Giving an initial population $\overline{\mathbf{u}}_{0}$, the dynamics is
given by the linear difference equation or matrix recurrence of the
multiplicative process:%
\begin{equation}
\overline{\mathbf{u}}_{n}=\mathbf{L}\overline{\mathbf{u}}_{n-1}\text{, with
}\overline{\mathbf{u}}_{0}\text{ given.} \label{Recufinite}%
\end{equation}
In finite matrix models when individuals in the older class $u_{p-1}$ become
infertile there is no need to go further than a $p\times p$ matrix. Variants
of this scheme use Usher \cite{Usher} or Lefkovitch \cite{Lefkovitch} matrices
and occur when dealing with modeling, simulation, experiments and observation
data from actual populations.

Two references in the subject of structured population dynamics in finite
classes are the books \cite{Caswell, Cushing01}.

In 1972 Lloyd Demetrius introduced the idea of an infinite Leslie matrix
\cite{Demetrius71,Demetrius72}, a natural generalization of the standard
Leslie matrix described by a countable infinite number of age-classes.

The extension of the classical Leslie models to infinite matrices was partly
motivated by an interest in understanding the existence and evolution of
mortality plateaus. This concept refers to an age-specific survivorship
distribution characterized by a mortality abates with age at advanced ages,
see for example, \cite{Fletcher98,Vaupel98}.

The asymptotic properties of the dynamical systems associated with the
infinite Leslie matrix was studied by considering the matrix as a positive
linear operator on a partially ordered Banach space
\cite{Demetrius71,Demetrius72}.

The infinite Leslie matrix is given by%
\begin{equation}
\mathbf{L}=\left(
\begin{array}
[c]{ccccc}%
a_{0} & a_{1} & a_{2} & a_{3} & \cdots\\
b_{1} & 0 & 0 & 0 & \cdots\\
0 & b_{2} & 0 & 0 & \cdots\\
0 & 0 & b_{3} & 0 & \cdots\\
\vdots & \vdots & \vdots & \vdots & \ddots
\end{array}
\right)  , \label{LeslieMatrix}%
\end{equation}
with almost the same interpretation of the finite one. So, we use the same
notation. The entries of this matrix are%
\begin{equation}
L_{ij}=\delta_{i1}a_{j-1}+\delta_{ij+1}b_{j},\text{ with }i,j\in\mathfrak{%
\mathbb{Z}
}^{+}, \label{entries}%
\end{equation}
the sets of indexes are denumerable. The same terminology as \cite{Seneta}
could be used calling the set of Leslie matrices (with finite or denumerable
index sets) the set of countable Leslie matrices. Since this paper deals
essentially with denumerable matrices we adhere to the name \textit{infinite
Leslie matrices} when referring to this type of matrices avoiding possible
ambiguities resulting from different terminologies in the literature. Our
results apply easily to the particular case of finite matrices as we will see below.

By appealing to the spectral theory of positive operators, Demetrius
\cite{Demetrius74} proved (see Theorems 2.1, 2.2 and 2.3 of the cited paper)
that when $%
{\textstyle\sum}
a_{n}<\infty$, $a_{n}$ for infinitely many $n$ and $b_{n}\rightarrow0$ as
$n\rightarrow0$, the model has an essentially unique stationary
age-distribution. (Essentially unique means in this context unique up to a
constant factor.) The asymptotic properties of the age-distribution were
analyzed in a later paper \cite{Demetrius72} and results analogous to the
asymptotic properties of finite Leslie matrices were obtained. Gosselin and
Lebreton, in a recent paper \cite{Gosselin}, relaxed the original Demetrius
conditions to derive analogous results concerning existence and asymptotic
properties of a stationary age distribution.

This paper is concerned with finite time properties of the discrete dynamical
system given by a equation similar to (\ref{Recufinite}), where the process is
now determined by an infinite Leslie matrix, denoted $\mathbf{L}$. The
corresponding discrete dynamical system is%
\begin{equation}
\overline{\mathbf{u}}_{n}=\mathbf{L}\overline{\mathbf{u}}_{n-1}\text{, }n\in%
\mathbb{N}
\text{,} \label{LeslieRec}%
\end{equation}
where $%
\mathbb{N}
$ is the set of the non-negative integers, $\overline{\mathbf{u}}$ is now a
population vector with a denumerable number of components and $\overline
{\mathbf{u}}_{0}$ is the initial condition.

In the literature \cite{Demetrius71,Demetrius72,Demetrius74,Gosselin} it is
assumed that the coefficients of $\mathbf{L}$ satisfy the following two conditions:

\begin{enumerate}
\item \label{RealWorld}$a_{l}\geq0$, $\sup a_{l}=A<\infty,$ with $l\in%
\mathbb{N}
$.

\item $0<b_{j}\leq1$ with $j\in\mathfrak{%
\mathbb{Z}
}^{+}$.
\end{enumerate}

These two conditions are consistent with the behaviour of natural
age-structured population. The conditions are also consistent with
mathematical models of self-reproducing machines \cite{Neumann, Freitas},
self-reproducing computer programs \cite{Burger, Kosa, Yamamoto}, or both,
like in the survey \cite{Sniper}. In this paper we do not need to assume these
conditions, since our method depends only on the form of the entries of the
infinite Leslie matrix (\ref{entries}).

Equation (\ref{SolucLeslie}) entails the following characterization of
$\overline{\mathbf{u}}_{n}$%

\begin{equation}
\overline{\mathbf{u}}_{n}=\mathbf{L}^{n}\overline{\mathbf{u}}_{0}\text{, for
}n\in%
\mathbb{N}
\text{.} \label{SolucLeslie}%
\end{equation}

The main aim of this article is the complete characterization of
$\overline{\mathbf{u}}_{n}$ for $n\in%
\mathbb{N}
$.

Our analysis revolves around the notion of a matrix of formal power series
$\mathbf{G}\left(  z\right)  \in\mathcal{M}^{\infty}\left[  \left[  z\right]
\right]  $, where $\mathcal{M}^{\infty}\left[  \left[  z\right]  \right]  $ is
the set of denumerable matrices which entries are formal power series. The
entries $\left[  G_{ij}\left(  z\right)  \right]  _{i,j\in\mathfrak{%
\mathbb{Z}
}^{+}}$ are functions of the Leslie matrix and the formal indeterminate $z$.

We will exploit the theory of kneading determinants to establish the relation%
\begin{equation}
\mathbf{G}\left(  z\right)  =%
{\displaystyle\sum\limits_{n\geq0}}
\mathbf{L}^{n}z^{n}\text{.} \label{GenMat}%
\end{equation}

This relation, which is the main result of the paper, provides a direct method
for computing $\mathbf{L}^{n}$ and thereby determining the solution of
equation (\ref{SolucLeslie}), in the cases of finite or infinite Leslie matrices.

We call $\mathbf{G}\left(  z\right)  $ the generating matrix of the solutions
of the difference equation.

This paper is organized as follows. In section \ref{SEC2} we describe with
illustrative examples, the characterization of $\mathbf{G}\left(  z\right)  $.
Section \ref{ProofMain} establishes the main result, as given by equation
(\ref{GenMat}).

\section{\label{SEC2}Statement of results}

The Cartesian product of $k\ $repeated sets $A$ is denoted by $A^{k}$. The
Cartesian product is denoted $A^{\infty}$ for a denumerable Cartesian product
of repeated sets.

$\mathcal{M}^{p}=%
\mathbb{C}
^{p\times p}$ and $\mathcal{M}^{\infty}=%
\mathbb{C}
^{\infty\times\infty}$ denote respectively the sets of $p\times p$ and
denumerable matrices with complex entries. $%
\mathbb{C}
\left[  z\right]  $ and $%
\mathbb{C}
\left[  \left[  z\right]  \right]  $ denote respectively the sets of formal
polynomials and formal power series in the indeterminate $z$ with complex
coefficients. The sets of $p\times p$ and denumerable matrices having formal
power series as entries are denoted respectively by $\mathcal{M}^{p}\left[
\left[  z\right]  \right]  $ and $\mathcal{M}^{\infty}\left[  \left[
z\right]  \right]  $. The indeterminates $x$ and $z$ are respectively
associated with real and complex indeterminates when nothing else mentioned.

We use a simplified notation to identify products of the transition
probabilities $b_{k}$ of a Leslie matrix%
\begin{equation}
\mathcal{C}_{k_{i}}^{k_{f}}=\left\{
\begin{array}
[c]{l}%
{\displaystyle\prod\limits_{k=k_{i}}^{k_{f}}}
b_{k}\text{ if }k_{f}\geq k_{i},\medskip\\
1\text{ if }k_{f}=k_{i}-1,\medskip\\
0\text{ otherwise.}%
\end{array}
\right.  \label{NotationC}%
\end{equation}
The usual notation, see \cite{Gosselin}, for these products is
\[
l_{n}=%
{\textstyle\prod\limits_{i=1}^{n-1}}
b_{n}=\mathcal{C}_{1}^{n-1},\text{ with }l_{1}=1\text{.}%
\]
In our proofs the bottom index $k_{i}$ in general is not $1$. This is the
reason for a special notation $\mathcal{C}_{k_{i}}^{k_{f}}$.

As usual in the case of sums if $k_{f}<k_{i}$ we have%
\[
\sum_{k=k_{i}}^{k_{f}}b_{k}=0.
\]

The usual discrete Heaviside function on the set of the integers $%
\mathbb{Z}
$ is denoted as%
\[
h\left(  k\right)  =\left\{
\begin{array}
[c]{l}%
1\text{ if }k\geq0,\\
0\text{ otherwise.}%
\end{array}
\right.
\]

The quantity%
\begin{equation}
\Delta\left(  z\right)  =1-\sum_{n\geq0}a_{n}\mathcal{C}_{1}^{n}z^{n+1},
\label{kneadingno}%
\end{equation}
is defined operationally here and named the weighed kneading determinant of
the Leslie difference equation (\ref{LeslieRec}) with weights $C_{1}^{n}%
=l_{n}$. The meaning of this concept will be made precise in section
\ref{ProofMain} where a general and deeper definition will be presented. We
see in the sequence that the real root of $\Delta\left(  z\right)  $ will be
associated with the solution of Euler-Lotka equation. The Definition
\ref{KDETFORMAL} of section \ref{ProofMain} agrees naturally with
$\Delta\left(  z\right)  $, when computed for Leslie difference equations.

\begin{definition}
\label{kneadingij}We define a matrix of formal power series $\mathbf{G}\left(
z\right)  \in\mathcal{M}^{\infty}\left[  \left[  z\right]  \right]  $, with
entries $\left[  G_{ij}\left(  z\right)  \right]  _{i,j\in\mathfrak{%
\mathbb{Z}
}^{+}}$ given by%
\begin{equation}
G_{ij}\left(  z\right)  =\mathcal{C}_{j}^{i-1}z^{i-j}+\frac{\mathcal{C}%
_{1}^{i-1}\sum\limits_{n\geq0}a_{j+n-1}\mathcal{C}_{j}^{j+n-1}z^{n+i}}%
{\Delta\left(  z\right)  }. \label{MainG}%
\end{equation}

\end{definition}

The zero degree coefficient of $\Delta\left(  z\right)  $ is $1$. Consequently
$\Delta\left(  z\right)  $ is invertible and the quotient in (\ref{MainG})
above is well defined. This definition will be clarified in Lemma \ref{Lemma}.

The main theorem of this paper states that:

\begin{theorem}
\label{MainTheorem}Given a Leslie matrix $\mathbf{L}$ and the matrix
$\mathbf{G}\left(  z\right)  \in\mathcal{M}^{\infty}\left[  \left[  z\right]
\right]  $ we have%
\[
\mathbf{G}\left(  z\right)  =\sum\limits_{n\geq0}\mathbf{L}^{n}z^{n}.
\]

\end{theorem}

This theorem holds even if the coefficients $a_{l}$ and $b_{j}$ of
$\mathbf{L}$ are general complex numbers. In the population dynamics context
of this work one considers real non-negative coefficients.

To compute $\mathbf{L}^{n}$ we have to identify the coefficients of $z^{n}$ in
$\mathbf{G}\left(  z\right)  $. Consequently, we call $\mathbf{G}\left(
z\right)  $ the generating matrix of the solutions of the Leslie difference
equation (\ref{LeslieRec}).

With this Theorem it is straightforward\ to compute the powers of $\mathbf{L}$
and to understand the dynamics of the difference equation (\ref{LeslieRec})
for different instances of Leslie matrices.

\begin{example}
\label{Example1}We have a particular simple situation for $G_{11}\left(
z\right)  $%
\[
G_{11}\left(  z\right)  =1+\frac{\sum\limits_{n\geq0}a_{n}\mathcal{C}_{1}%
^{n}z^{n+1}}{\Delta\left(  z\right)  },
\]
we have $\sum\limits_{n\geq0}a_{n}\mathcal{C}_{1}^{n}z^{n+1}=1-\Delta\left(
z\right)  $, hence%
\[
=1+\frac{1-\Delta\left(  z\right)  }{\Delta\left(  z\right)  }=\frac{1}%
{\Delta\left(  z\right)  }.
\]
The generating function for $L_{11}^{n}$ (the entries $\left(  1,1\right)  $
of the powers of $\mathbf{L}$) is the inverse of the weighed kneading
determinant. This agrees with the direct computations of the matrix elements
of $\mathbf{L}^{n}$ giving $L_{11}^{0}=1$, $L_{11}=a_{0}$, $L_{11}^{2}%
=a_{0}^{2}+a_{1}b_{1}$, $L_{11}^{3}=a_{0}^{3}+2a_{0}a_{1}b_{1}+a_{2}b_{1}%
b_{2}$, etc.
\end{example}

The equation
\begin{equation}
1-\sum_{n\geq0}a_{n}\mathcal{C}_{1}^{n}\left(  \frac{1}{\rho}\right)
^{n+1}=0, \label{Euler-Lotka2}%
\end{equation}
is the well known Euler-Lotka equation \cite{Gosselin} for the leading
eigenvalue, $\rho>0$, for an infinite Leslie matrix. If we identify
$z=\rho^{-1}$ in (\ref{Euler-Lotka2}) we get $\Delta\left(  z\right)  =0$.
Consequently, the weighed kneading determinant and its roots are very
important in terms of asymptotic behavior of the solutions.

It is not our purpose to study a plethora of concrete examples that can be
readily obtained from the vast literature on this subject, see for example
\cite{Caswell,Cushing01}. We present two case studies only as examples of the
computations involved. The heart of the matter is that any Leslie difference
equation, finite or infinite, can be solved using kneading theory. In the
finite case, the classic method of diagonalization to find powers of matrices
works perfectly. The kneading determinant method is a new technique that as
the advantage, in some cases, to be computationally lighter.

\begin{example}
\label{Finite}Finite Leslie model.

A possible, very simplified, Leslie matrix for semelparous population with
three age classes \cite{Cushing} is
\[
\mathbf{L}=\left(
\begin{array}
[c]{ccc}%
0 & 0 & a\\
b_{1} & 0 & 0\\
0 & b_{2} & 0
\end{array}
\right)  ,
\]
where $a$ is the average fertility rate of each female, $b_{1}$ is the
survival rate of the newborn generation and $b_{2}$ is the survival rate of
the juveniles. See a more general discussion in \cite{Cushing} with an
extensive related bibliography.

The weighed kneading determinant is%
\[
\Delta\left(  z\right)  =1-ab_{1}b_{2}z^{3}.
\]
The generating matrix for this model is%
\[
\mathbf{G}\left(  z\right)  =\frac{1}{1-ab_{1}b_{2}z^{3}}\left(
\begin{array}
[c]{ccc}%
1 & ab_{2}z^{2} & az\\
b_{1}z & 1 & ab_{1}z^{2}\\
b_{1}b_{2}z^{2} & b_{2}z & 1
\end{array}
\right)  ,
\]
with an easy expansion in power series giving%
\[
\mathbf{L}^{3n}\mathbf{=}\left(  ab_{1}b_{2}\right)  ^{n}\left(
\begin{array}
[c]{ccc}%
1 & 0 & 0\\
0 & 1 & 0\\
0 & 0 & 1
\end{array}
\right)  ,\text{ }\mathbf{L}^{3n+1}\mathbf{=}\left(  ab_{1}b_{2}\right)
^{n}\left(
\begin{array}
[c]{ccc}%
0 & 0 & a\\
b_{1} & 0 & 0\\
0 & b_{2} & 0
\end{array}
\right)  ,
\]
and
\[
\mathbf{L}^{3n+2}\mathbf{=}\left(  ab_{1}b_{2}\right)  ^{n}\left(
\begin{array}
[c]{ccc}%
0 & ab_{2} & 0\\
0 & 0 & ab_{1}\\
b_{1}b_{2} & 0 & 0
\end{array}
\right)
\]
The general semelparous case with $p$ age classes can be obtained using
similar reasonings.

The Euler-Lotka equation has solution $z_{0}=\frac{1}{\sqrt[3]{ab_{1}b_{2}}}$
or $\rho=\sqrt[3]{ab_{1}b_{2}}$. In the case $ab_{1}b_{2}=1$ the population is
stable with yearly oscillations of period $3$.
\end{example}

The possibility of infinite Leslie matrices can occur when the population does
age very slowly or when we consider each member of the population as a colony
of insects like ants or termites. Theoretically these colonies can last
forever. Each colony can reproduce founding nearby colonies. The scale of the
reproduction is very slow but each colony has high survivorship rate. So, the
reproducing coefficients $a_{l}$ are very small and the transition
probabilities $b_{j}$ near $1$. Despite these biological considerations, the
relative size of the coefficients $a_{l}$ and $b_{j}$ is immaterial to the
computation of the finite time solutions, as we can see in the next example.

\begin{example}
\label{Ex1}Infinite Leslie model.

As an ideal experiment, which is a very simple generalization of the models
presented in \cite{Gosselin}, we consider populations were the reproduction
varies geometrically with age and with a mortality rate constant. Geometric
birth rates are used since the formal power series are readily obtained
exhibiting the internal machinery of this procedure. Therefore, we have%
\[
\mathbf{L}=\left(
\begin{array}
[c]{ccccc}%
a & a\chi & a\chi^{2} & a\chi^{3} & \cdots\\
b & 0 & 0 & 0 & \cdots\\
0 & b & 0 & 0 & \cdots\\
0 & 0 & b & 0 & \cdots\\
\vdots & \vdots & \vdots & \vdots & \ddots
\end{array}
\right)  .
\]
The case $\chi=1$ was treated in\cite{Gosselin}. We compute the weighed
kneading determinant%
\[
\Delta\left(  z\right)  =1-a\sum_{n\geq0}\mathcal{C}_{1}^{n}\chi^{n}%
z^{n+1}=1-a\sum_{n\geq0}b^{n}\chi^{n}z^{n+1}=\frac{1-\left(  a+b\chi\right)
z}{1-b\chi z}.
\]
The generating functions are
\begin{align*}
G_{ij}\left(  z\right)   &  =\mathcal{C}_{j}^{i-1}z^{i-j}+\frac{\mathcal{C}%
_{1}^{i-1}\sum\limits_{n\geq0}a_{j+n-1}\mathcal{C}_{j}^{j+n-1}z^{n+i}}%
{\Delta\left(  z\right)  }\\
&  =b^{i-j}z^{i-j}h\left(  i-j\right)  +ab^{i-1}\chi^{j-1}\sum_{n\geq
i}\left(  a+b\chi\right)  ^{n-i}z^{n},
\end{align*}
where $h$ is the discrete Heaviside function.

To simplify the notation, the generating matrix $\mathbf{G}\left(  z\right)
=\left[  G_{ij}\left(  z\right)  \right]  _{i,j=1,2,\ldots}$ can be written as%
\[
\mathbf{G}\left(  z\right)  =\left(
\begin{array}
[c]{ccccc}%
1+g_{11}\left(  z\right)  & g_{12}\left(  z\right)  & g_{13}\left(  z\right)
& g_{14}\left(  z\right)  & \cdots\\
bz+g_{21}\left(  z\right)  & 1+g_{22}\left(  z\right)  & g_{23}\left(
z\right)  & g_{24}\left(  z\right)  & \cdots\\
b^{2}z^{2}+g_{31}\left(  z\right)  & bz+g_{32}\left(  z\right)  &
1+g_{33}\left(  z\right)  & g_{34}\left(  z\right)  & \cdots\\
b^{3}z^{3}+g_{41}\left(  z\right)  & b^{2}z^{2}+g_{42}\left(  z\right)  &
bz+g_{43}\left(  z\right)  & 1+g_{44}\left(  z\right)  & \cdots\\
\vdots & \vdots & \vdots & \vdots & \ddots
\end{array}
\right)  ,
\]
where, in this case%
\[
g_{ij}\left(  z\right)  =ab^{i-1}\chi^{j-1}\sum_{n\geq i}\left(
a+b\chi\right)  ^{n-i}z^{n}.
\]
The above expression for the entries of $\mathbf{G}\left(  z\right)  $ solves
explicitly the difference equation (\ref{LeslieRec}) and gives the entries of
$\mathbf{L}^{n}$.

The Euler-Lotka equation%
\[
\frac{1-\left(  a+b\chi\right)  \rho^{-1}}{1-b\chi\rho^{-1}}=0
\]
has solution $\rho^{-1}=z=\frac{1}{a+b\chi}$. Therefore, the leading
eigenvalue is $\rho=a+b\chi$ and the replacement fertility (fertility rate
that keeps stable the population average) is $a=1-b\chi$.
\end{example}

\section{\label{ProofMain}Proof of the main result}

The proof of the main Theorem \ref{MainTheorem} is based on the concept of
kneading determinant of a pair of linear endomorphisms which difference has
finite rank. A concept \ that roots its origin in one dimensional dynamics
\cite{Milnor}. In the context of linear algebra an analogous definition of
kneading determinant using formal power series can be obtained, see
\cite{jalves1}. The process has similarities to the original definition for
discrete dynamics of the interval, which justifies the use of the terminology
borrowed from dynamical systems.

\begin{remark}
\label{Note}The matrix $\mathbf{L}$ can be decomposed in a sum of two matrices
$\mathbf{L}=\mathcal{R}+\mathbf{X}$ given by%
\[
\mathbf{X}=\left(
\begin{array}
[c]{ccccc}%
0 & 0 & 0 & 0 & \cdots\\
b_{1} & 0 & 0 & 0 & \cdots\\
0 & b_{2} & 0 & 0 & \cdots\\
0 & 0 & b_{3} & 0 & \cdots\\
\vdots & \vdots & \vdots & \vdots & \ddots
\end{array}
\right)  \text{ and }\mathcal{R}=\left(
\begin{array}
[c]{ccccc}%
a_{0} & a_{1} & a_{2} & a_{3} & \cdots\\
0 & 0 & 0 & 0 & \cdots\\
0 & 0 & 0 & 0 & \cdots\\
0 & 0 & 0 & 0 & \cdots\\
\vdots & \vdots & \vdots & \vdots & \ddots
\end{array}
\right)  ,
\]
please note that $\mathcal{R}\mathbf{=L}-\mathbf{X}$ is the matrix
representation of an endomorphism in $%
\mathbb{C}
^{\infty}$ with rank $1$. This decomposition is classic and was noticed in
\cite{Caswell, Cushing, Usher}, in different frameworks.
\end{remark}

\begin{definition}
Consider the standard basis of $%
\mathbb{C}
^{\infty}$, the linear space of infinite sequences over the field of complex
numbers. The vectors of this basis are the columns of the identity infinite
matrix $\left(  C_{l}\right)  _{l=1,2,\ldots}$. On the other side $R_{\alpha}$
is the $\alpha$-th row of the infinite identity matrix, or the $\alpha$ vector
of the standard basis of the dual space of $%
\mathbb{C}
^{\infty}$.
\end{definition}

We denote by $R$ the first row of $\mathbf{L}$,%
\[
R=R_{1}\mathcal{R}=\left(
\begin{array}
[c]{cccccc}%
a_{0} & a_{1} & a_{2} & \cdots & a_{l} & \cdots
\end{array}
\right)  .
\]

The matrix $\mathbf{X}$ has entries%
\begin{equation}
X_{ij}=\delta_{ij+1}b_{j}, \label{Xcomp}%
\end{equation}
where $\delta_{ij}$ is the usual delta Kronecker symbol, therefore
$\mathbf{X}$ has very easy powers $\mathbf{X}^{n}$, with $n\in%
\mathbb{N}
$.

For $n\in%
\mathbb{N}
$ the powers of $\mathbf{X}$ have entries
\[
X_{ij}^{n}=\delta_{ij+n}%
{\displaystyle\prod\limits_{k=0}^{n-1}}
b_{j+k}.
\]
This is a direct consequence of the equality (\ref{Xcomp}). Naturally when
$\mathbf{L}$ is finite, $\mathbf{X}$ is nilpotent.

\begin{definition}
The weighed kneading invariant (a particular case of the kneading matrix of
section 3 of \cite{jalves2}) is the formal power series in $%
\mathbb{C}
\left[  \left[  z\right]  \right]  $ associated with the Leslie difference
equation (\ref{LeslieRec}) and given by%
\begin{equation}
M\left(  z\right)  =\sum_{n\geq0}R\mathbf{X}^{n}C_{1}z^{n}\text{.}
\label{Knead}%
\end{equation}
The weights appear as the products of the $b_{j}$ of the Leslie matrix and are
not present at the sections one and two of \cite{jalves2}.
\end{definition}

\begin{proposition}
The weighed kneading invariant $M\left(  z\right)  \in%
\mathbb{C}
\left[  \left[  z\right]  \right]  $ of the Leslie difference equation
(\ref{LeslieRec}) is
\[
M\left(  z\right)  =\sum_{n\geq0}a_{n}\mathcal{C}_{1}^{n}z^{n}\text{.}%
\]

\end{proposition}

\begin{pf}
Note that $\mathbf{X}^{n}C_{1}$ is the first column of $\mathbf{X}^{n}$, with
its entries given by%
\[
X_{i1}^{n}=\delta_{i1+n}%
{\displaystyle\prod\limits_{k=0}^{n-1}}
b_{1+k}=\delta_{i1+n}%
{\displaystyle\prod\limits_{k=1}^{n}}
b_{k}=\delta_{i1+n}\mathcal{C}_{1}^{n}.
\]
Replacing $X_{i1}^{n}$ in (\ref{Knead}) we get the desired result.
\end{pf}

We provide here the formal definition of weighed kneading determinant of a
Leslie difference equation defined operationally for Leslie difference
equations in equation (\ref{kneadingno}).

\begin{definition}
\label{KDETFORMAL}Giving the weighed kneading invariant $M\left(  z\right)  ,$
the weighed kneading determinant $\Delta\left(  z\right)  \in%
\mathbb{C}
\left[  \left[  z\right]  \right]  $ (see \cite{jalves2}) associated to
$M\left(  z\right)  $ is
\[
\Delta\left(  z\right)  =1-zM\left(  z\right)  .
\]

\end{definition}

The use of the concept \emph{determinant} is inherited from the higher
dimensional case of infinite Fibonacci difference equations \cite{jalves2}.

\begin{remark}
When we compute the weighed kneading determinant with the above definition,
the result agrees with (\ref{kneadingno})\ used previously%
\[
\Delta\left(  z\right)  =1-zM\left(  z\right)  =1-\sum_{n\geq0}a_{n}%
\mathcal{C}_{1}^{n}z^{n+1}.
\]

\end{remark}

\begin{definition}
The extended weighed kneading matrices for $i,j\in\mathfrak{%
\mathbb{Z}
}^{+}$ (again from section 3 of \cite{jalves2}) are defined such that%
\begin{equation}
\mathbf{M}_{ij}\left(  z\right)  =\left(
\begin{array}
[c]{cc}%
{\displaystyle\sum\limits_{n\geq0}}
R\mathbf{X}^{n}C_{1}z^{n} &
{\displaystyle\sum\limits_{n\geq0}}
R\mathbf{X}^{n}C_{j}z^{n}\\%
{\displaystyle\sum\limits_{n\geq0}}
R_{i}\mathbf{X}^{n}C_{1}z^{n} &
{\displaystyle\sum\limits_{n\geq0}}
R_{i}\mathbf{X}^{n}C_{j}z^{n}%
\end{array}
\right)  , \label{Extended}%
\end{equation}
with $\mathbf{M}_{ij}\left(  z\right)  \in\mathcal{M}^{2}\left[  \left[
z\right]  \right]  $.
\end{definition}

\begin{proposition}
The extended weighed kneading matrices are%
\[
\mathbf{M}_{ij}\left(  z\right)  =\left(
\begin{array}
[c]{cc}%
\sum\limits_{n\geq0}a_{n}\mathcal{C}_{1}^{n}z^{n} & \sum\limits_{n\geq
0}a_{j+n-1}\mathcal{C}_{j}^{j+n-1}z^{n}\\
\mathcal{C}_{1}^{i-1}z^{i-1} & \mathcal{C}_{j}^{i-1}z^{i-j}%
\end{array}
\right)  .
\]

\end{proposition}

\begin{pf}
Recall that%
\[
X_{ij}^{n}=\delta_{ij+n}%
{\displaystyle\prod\limits_{k=0}^{n-1}}
b_{j+k}=\delta_{ij+n}%
{\displaystyle\prod\limits_{k=j}^{j+n-1}}
b_{k}=\delta_{ij+n}\mathcal{C}_{j}^{j+n-1},
\]
so%
\[%
{\displaystyle\sum\limits_{n\geq0}}
R\mathbf{X}^{n}C_{j}z^{n}=\sum_{n\geq0}a_{j+n-1}\mathcal{C}_{j}^{j+n-1}%
z^{n}\text{.}%
\]
The first row of $\mathbf{M}_{ij}\left(  z\right)  $ is done. Now, we have to
compute the last row of $\mathbf{M}_{ij}\left(  z\right)  $. We note that
there is only one entry different from $0$ at the $j$ column of $\mathbf{X}%
^{n}$ and that element is at line $i$ which is greater than $j$, so%
\[%
{\displaystyle\sum\limits_{n\geq0}}
R_{i}\mathbf{X}^{n}C_{j}z^{n}=\mathcal{C}_{j}^{i-1}z^{i-j},
\]
as desired.
\end{pf}

\begin{definition}
The extended weighed kneading determinant of $\mathbf{M}_{ij}\left(  z\right)
$ is
\[
\Delta_{ij}\left(  z\right)  =\det\left(  \mathbf{I}-z\mathbf{M}_{ij}\left(
z\right)  \right)  \text{,}%
\]
where $\mathbf{I}$ denotes the identity matrix. It is a formal power series in
$%
\mathbb{C}
\left[  \left[  z\right]  \right]  .$
\end{definition}

The next Lemma contains all the relevant details necessary to prove the main
Theorem \textbf{\ref{MainTheorem}}.

\begin{lemma}
\label{Lemma}The generating functions (see \cite{jalves1} and \cite{jalves2})
$G_{ij}\left(  z\right)  \in%
\mathbb{C}
\left[  \left[  z\right]  \right]  $ and $i,j\in\mathfrak{%
\mathbb{Z}
}^{+}$ $\ $for the solutions of (\ref{LeslieRec}) are%
\[
G_{ij}\left(  z\right)  =\frac{1}{z}\left(  1-\frac{\Delta_{ij}\left(
z\right)  }{\Delta\left(  z\right)  }\right)  ,
\]
this generating functions are formal power series and have as coefficients of
$z^{n}$ the matrix elements $L_{ij}^{n}\,\ $(the entry $\left(  i,j\right)  $
of the matrix $\mathbf{L}^{n}$). The entries $G_{ij}\left(  z\right)  $ are%
\[
G_{ij}\left(  z\right)  =\mathcal{C}_{j}^{i-1}z^{i-j}+\frac{\mathcal{C}%
_{1}^{i-1}\sum\limits_{n\geq0}a_{j+n-1}\mathcal{C}_{j}^{j+n-1}z^{n+i}}%
{\Delta\left(  z\right)  }.
\]

\end{lemma}

\begin{pf}
The proof of this Lemma follows the proof of Theorem 2.2 in \cite{jalves2} if
we associate $\mathbf{X}$ with the matrix representation of the endomorphism
$\varphi$ and $\mathbf{L}$ with the matrix representation of the endomorphism
$\psi$. From Remark \ref{Note}, $\psi-\varphi$ has finite rank and all the
conclusions of \cite{jalves1} and \cite{jalves2} apply to $\psi^{n}$ with
matrix representation $\mathbf{L}^{n}$, as desired.

To prove the second statement we compute explicitly the extended weighed
kneading determinants. We have%
\[
\Delta_{ij}\left(  z\right)  =\det\left(
\begin{array}
[c]{cc}%
1-z\sum\limits_{n\geq0}a_{n}\mathcal{C}_{1}^{n}z^{n} & -z\sum\limits_{n\geq
0}a_{j+n-1}\mathcal{C}_{j}^{j+n-1}z^{n}\\
-z\mathcal{C}_{1}^{i-1}z^{i-1} & 1-z\mathcal{C}_{j}^{i-1}z^{i-j}%
\end{array}
\right)  ,
\]
which is%
\[
\Delta_{ij}\left(  z\right)  =\Delta\left(  z\right)  \left(  1-\mathcal{C}%
_{j}^{i-1}z^{i-j+1}\right)  -\mathcal{C}_{1}^{i-1}%
{\textstyle\sum\limits_{n\geq0}}
a_{j+n-1}\mathcal{C}_{j}^{j+n-1}z^{n+i+1}.
\]
From
\begin{align*}
zG_{ij}\left(  z\right)   &  =1-\frac{\Delta_{ij}\left(  z\right)  }%
{\Delta\left(  z\right)  }\\
&  =1-\frac{\Delta\left(  z\right)  \left(  1-\mathcal{C}_{j}^{i-1}%
z^{i-j+1}\right)  -\mathcal{C}_{1}^{i-1}%
{\textstyle\sum\limits_{n\geq0}}
a_{j+n-1}\mathcal{C}_{j}^{j+n-1}z^{n+i+1}}{\Delta\left(  z\right)  }\\
&  =\mathcal{C}_{j}^{i-1}z^{i-j+1}+\frac{\mathcal{C}_{1}^{i-1}\sum
\limits_{n\geq0}a_{j+n-1}\mathcal{C}_{j}^{j+n-1}z^{n+i+1}}{\Delta\left(
z\right)  },
\end{align*}
we obtain the generating functions%
\[
G_{ij}\left(  z\right)  =\mathcal{C}_{j}^{i-1}z^{i-j}+\frac{\mathcal{C}%
_{1}^{i-1}\sum\limits_{n\geq0}a_{j+n-1}\mathcal{C}_{j}^{j+n-1}z^{n+i}}%
{\Delta\left(  z\right)  },
\]
as desired.
\end{pf}

The matrix $\mathbf{G}\in\mathcal{M}^{\infty}\left[  \left[  z\right]
\right]  $ with entries $G_{ij}\left(  z\right)  $, is the generating matrix
of the solutions of the Leslie difference equation (\ref{LeslieRec}).

We have all the ingredients to prove the Main Theorem \ref{MainTheorem}. The
result follows directly from Lemma \ref{Lemma} and the definition of
generating function, i.e.,%
\[
G_{ij}\left(  z\right)  =\sum\limits_{n\geq0}L_{ij}^{n}z^{n}\text{, }i,j\in%
\mathbb{Z}
^{+}\text{.}%
\]

In our next article we use the main theorem of this work to study the
asymptotic properties of models with infinite Leslie matrices.

\begin{Conclu}
We point out that with generating functions for $\mathbf{L}^{n}$ explicitly
given, one gets the solutions of this type of difference equations under very
mild restrictions, which is a clear benefit of working with formal power series.
\end{Conclu}

\noindent\textbf{Acknowledgement }The contribution of the anonymous referees
improved a great deal our final version of the paper, we thank both for the
relevant comments and suggestions. We also thank the invaluable discussions we
sustained with Lloyd Demetrius and the ideas of Jim Cushing.

\end{document}